\documentclass[twocolumn,preprintnumbers,amsmath,amssymb,aps,prc,groupedaddress,floatfix]{revtex4-1}

\usepackage{graphicx}
\usepackage{dcolumn}
\usepackage{nicefrac}
\usepackage{bm}
\usepackage{url}

\usepackage[T1]{fontenc}
\usepackage[percent]{overpic}
\usepackage{tikz}
\usetikzlibrary{arrows}
\usetikzlibrary{positioning}

\usepackage{sansmath}

\begin{document}

\newcommand{\element}[2]{$^{#1}$\textrm{#2}}

\title{Perturbed angular distributions with LaBr$_3$ detectors:
  the $\bm{g}$ factor of the first $\bm{10^+}$ state in $^{110}$Cd revisited}

\author{T.~J.~Gray}
\author{A.~E.~Stuchbery}
\author{M.~W.~Reed}
\author{A.~Akber}
\author{B.~J.~Coombes}
\author{J.~T.~H.~Dowie}
\author{T.~K.~Eriksen}
\author{M.~S.~M.~Gerathy}
\author{T.~Kib\'edi}
\author{G.~J.~Lane}
\author{A.~J.~Mitchell}
\author{T.~Palazzo}
\author{T.~Tornyi}
\affiliation{Department of Nuclear Physics, Research School of Physics
  and Engineering, The Australian National University, Canberra, ACT 2601,
  Australia}

\date{\today}

\begin{abstract}
  The Time Differential
  Perturbed Angular Distribution technique with LaBr$_3$ detectors has
  been applied to the $I^\pi = \frac{11}{2}^-$ isomeric state ($E_x = 846$
  keV, $\tau=107$~ns) in $^{107}$Cd, which was populated and recoil-implanted
  into a gadolinium host following the
  $^{98}$Mo($^{12}$C, $3n$)$^{107}$Cd reaction. The static hyperfine
  field strength of Cd recoil implanted into gadolinium was thus measured,
  together with the fraction of nuclei implanted into field-free
  sites, under similar conditions as pertained for a previous implantation perturbed angular distribution
  $g$-factor measurement on the $I^\pi = 10^+$ state in
  $^{110}$Cd. The $^{110}$Cd $g(10^+)$ value was
  thereby re-evaluated, bringing it into agreement with the value
  expected for a seniority-two
  $\nu h_{\nicefrac{11}{2}}$ configuration.
\end{abstract}

\maketitle
\section{Introduction}\label{sect:intro}
The recent development of Lanthanum Bromide (LaBr$_3$) scintillator
detectors provides an opportunity to perform perturbed angular
distribution $g$-factor measurements under new experimental
conditions.  In such measurements, the time-dependent spin-rotation of
a nuclear state subjected to a known magnetic field can be observed to
deduce the nuclear $g$ factor.  One limitation of time-differential techniques is the maximum
frequency that can be resolved by the experimental system.  High-purity germanium (HPGe)
detectors are commonly used for in-beam spectroscopy due to their
excellent energy resolution.  However, the time resolution of HPGe
detectors (${\sim}10$~ns) is insufficient for many
time-differential measurements~\cite{Mihailescu2007, Crespi2010}. In
particular, HPGe detectors cannot be used in cases that use strong
hyperfine magnetic fields with resultant high precession frequencies
corresponding to periods $\lesssim 10$~ns.

LaBr$_3$ detectors have excellent time resolution ({${\sim}300$~ps} is
readily achievable~\cite{Iltis2006, Regis2010,
Mason2013, Alharbi2013, Roberts2014, Marginean2010,
Regis2009, Zhu2011}), and
energy resolution much superior to other commonly used scintillators
such as NaI and BaF$_2$, making
them an excellent choice for fast in-beam Time Differential
  Perturbed Angular Distribution (TDPAD) measurements~\cite{Fraile2017}.  Thus the new opportunities
for application of LaBr$_3$ detectors include the measurement of \mbox{${g}$
factors} of shorter-lived nuclear states in known, intense, static hyperfine magnetic
fields in magnetic hosts, satisfying the condition $\tau \gtrsim
T_L =
\pi/\omega_L$, where $\tau$ is the nuclear mean life, and
$T_L(\omega_L)$ is the Larmor period (frequency)~\cite{Cerny1974}.
Cases where the period $T_L$ is of the order of a few nanoseconds become
accessible for in-beam measurement. The measurements typically have
the target backed by a ferromagnetic foil into which the nuclei of
interest are recoil implanted.  Available ferromagnetic hosts
include iron, nickel, cobalt, and gadolinium. Gadolinium is an
advantageous ferromagnetic host due to its higher $Z$, which allows nuclei with
$Z\lesssim 60$ to be created in-beam at energies below the Coulomb
barrier of the beam on gadolinium, resulting in cleaner $\gamma$-ray spectra.

As a first application of LaBr$_3$ detectors to in-beam TDPAD
techniques, the hyperfine field of Cd implanted into gadolinium has been
investigated.  The motivation was to revisit the $g$-factor measurement
on the \mbox{$I^\pi=10^{+}$} state in $^{110}$Cd~\cite{Regan1995}.  Even though the \mbox{$I^\pi=10^+$} itself is too short lived
(\mbox{$\tau \sim 1$~ns~\cite{Piiparinen1993, Juutinen1994, Kostov1998,
  Harissopulos2001}}) to apply  the TDPAD method directly, in-beam
TDPAD measurements can determine the effective hyperfine
field at Cd nuclei implanted into gadolinium hosts.  Additionally, the original measurement was time-integral, and
reported \mbox{$g(10^+)=-0.09(3)$}, at least a factor of two smaller than
\mbox{$g\approx-0.2$ to $-0.3$} that would be expected for a rather pure $(h_{\nicefrac{11}{2}})^2$ neutron
configuration~\cite{Regan1995}. In contrast, recent laser spectroscopy
on odd-mass Cd
isotopes shows a sequence of low-lying $\nu h_{\nicefrac{11}{2}}$ states with $g \sim
-0.2$~\cite{Frommgen2015, Yordanov2013}, and the $I^\pi=10^+$ state in
$^{110}$Cd is expected to have a
similar $g$~factor.

The
reaction used in Ref.~\cite{Regan1995} was $^{100}$Mo($^{13}$C,
$3n$)$^{110}$Cd. An attempt was made in
Ref.~\cite{Regan1995} to calibrate the effective hyperfine field,
$B_{\rm hf}$, of Cd in gadolinium after recoil implantation. The
$^{100}$Mo($^{12}$C, $5n$)$^{107}$Cd reaction on the same target populated a convenient
$I^\pi = \frac{11}{2}^-$, $\tau=107$~ns isomer in $^{107}$Cd, with a
known $g$ factor~\cite{Bertschat1974}. The attempt was
unsuccessful, however, because the expected precession period $T_L
\approx 12$~ns could not be resolved by the HPGe detectors. We
have revisited this measurement using LaBr$_3$ detectors and the
$^{98}$Mo($^{12}$C, $3n$)$^{107}$Cd reaction, under similar conditions
to the $^{110}$Cd $g$-factor measurement.

The motivation for the experiment was therefore threefold: first, to gain
experience using LaBr$_3$ detectors in the context of in-beam TDPAD
techniques; second, to evaluate gadolinium as a
ferromagnetic host for in-beam $g$-factor measurements; and third, to
understand why the measured $g$ factor in $^{110}$Cd was
lower than expected.
\section{Experiment}\label{sect:expt}
\begin{figure}
  \includegraphics[width=0.45\textwidth]{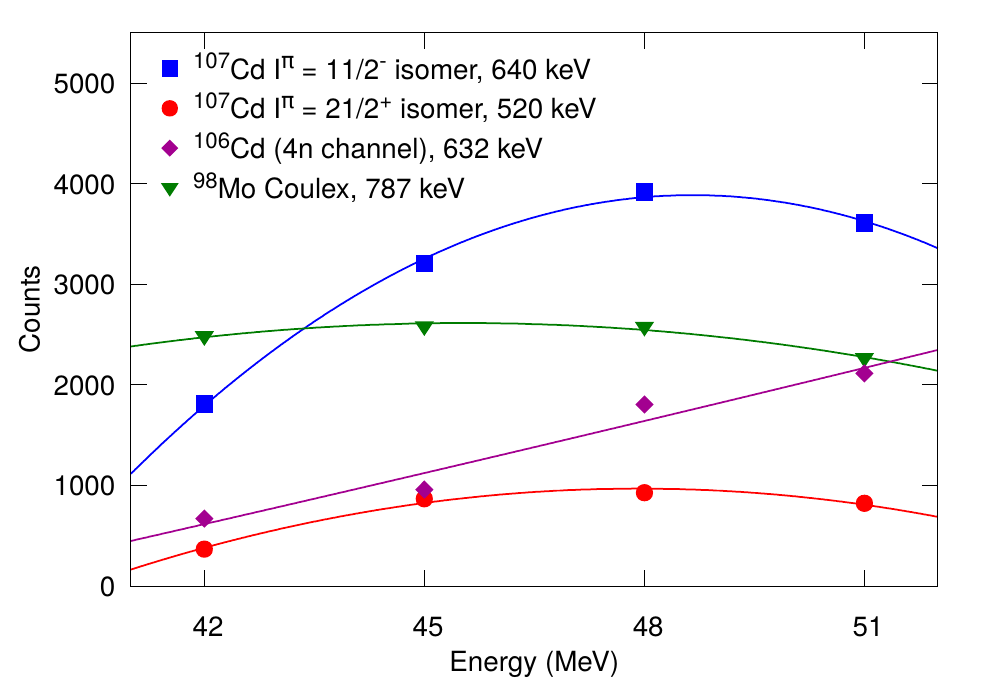}
  \caption{Excitation function for $\gamma$-ray
    energies from different channels for $^{12 }$C beams on a $^{98}$Mo target with
    fits to guide the eye.}
  \label{fig:excitation}
\end{figure}

The experiment used a 48-MeV $^{12}$C beam delivered by the ANU 14UD Pelletron
accelerator. Figure~\ref{fig:excitation} shows the excitation
functions used to select the most favourable beam energy. The beam was pulsed in
bunches of ${\sim} 2$~ns full width at half maximum (FWHM) separated by 963~ns.  A two-layer target
consisting of 97.7\% $^{98}$Mo 280-$\mu$g/cm$^2$ thick evaporated onto
a 99.9\% pure natural gadolinium foil was used.  This foil was rolled to a thickness of 3.94~mg/cm$^2$, before being annealed at
${\sim} 770^\circ$~C in vacuum for 20 minutes. The reaction $^{98}$Mo($^{12}$C,
$3n$)$^{107}$Cd occurs in the first target layer, and the $^{107}$Cd
nuclei recoil and stop in the second (gadolinium) layer.
The ANU hyperfine spectrometer~\cite{Stuchbery2017} was used for the experiment.  The
target was cooled to ${\sim}$ 6~K, and an external field of ${\sim}$ 0.1~T
applied in the vertical direction to polarize the ferromagnetic gadolinium layer.
Four LaBr$_3$ detectors placed in the horizontal plane at
$\pm 45^\circ$ and $\pm135^\circ$ with respect to the beam direction were
located 79 mm from the target. The LaBr$_3$ crystals were 38 mm in
diameter and 51 mm long.  Mu-metal shielding, both
around the target chamber and each detector, served to eliminate the
stray field from the electromagnet. The magnetic field direction was
reversed periodically to reduce systematic errors.

A single Ortec 567 Time to Amplitude Converter on the $1$ $\mu$s range
collected
timing information for all four detectors with respect to the beam pulse.

The reaction excited the $^{107}$Cd nucleus, populating the 846-keV,
$I^\pi=\frac{11}{2}^-$ state of interest with mean life $\tau~=~107(3)$~ns~\cite{Regan1995},
and $g$ factor $g=-0.189(2)$~\cite{Raghavan1989}. For a hyperfine field
strength of $B_{\rm hf}=-34.0(7)$~T~\cite{Krane1983, Forker1973} the expected precession period is $T_L =
\pi/\omega_L \approx 12$~ns.

A HPGe detector was also present for a high energy-resolution monitor
at $\theta \approx -90^\circ$ with respect to the beam axis. To
measure angular distributions, the
LaBr$_3$ detectors at negative angles were removed, and the angle of
the HPGe detector varied to \mbox{$\lvert \theta \rvert=0^\circ, 25^\circ, 45^\circ,
65^\circ,$ and $90^\circ$}.

A post-experiment inspection
of the target by eye showed no obvious physical damage, however
build-up of carbon on the back of the foil was observed.
\section{Results}\label{sect:results}
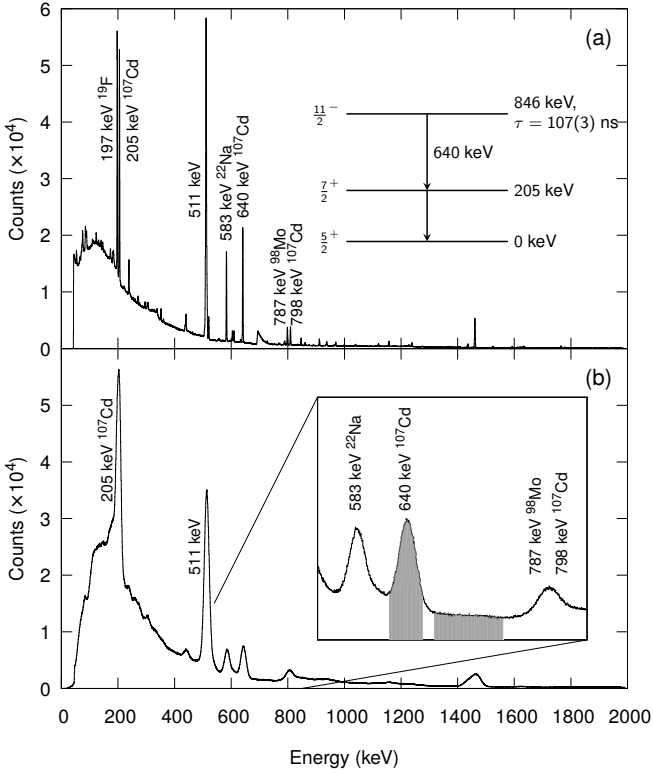
\begin{figure}
  \begin{overpic}[width=0.5\textwidth]{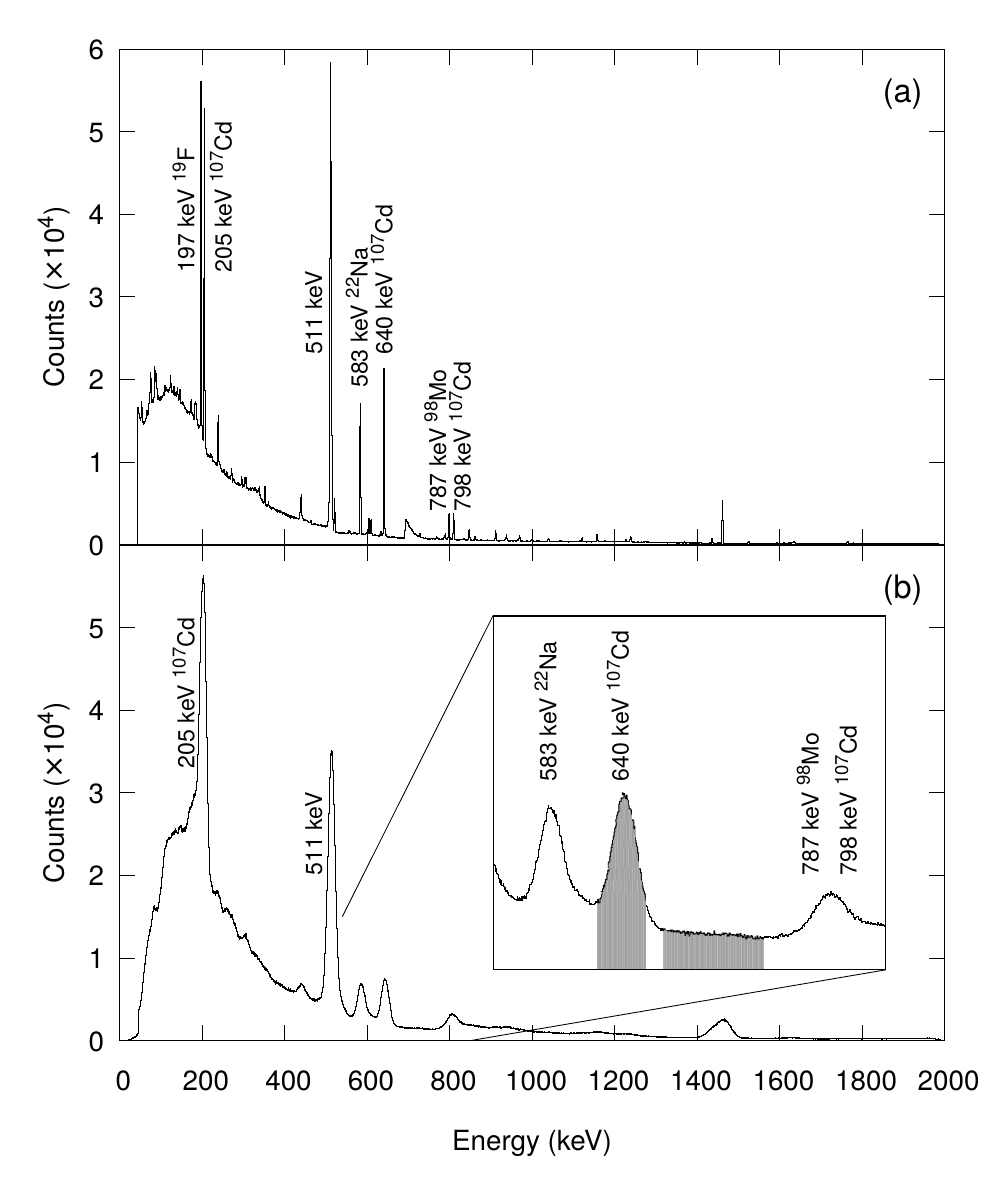}
    \put(38, 65){
      \resizebox{4.5cm}{!}{
        \fontfamily{phv}\selectfont
        \begin{sansmath}
    \begin{tikzpicture}[
        level/.style={thick},
        transition/.style={thick,->,>=stealth},
        scale=3]
      \draw[level] (1cm,-0.5em)node[right]{$0$ keV} -- (0cm, -0.5em) node[left]{$\frac{5}{2}^+$};
      \draw[level] (1cm,0.5em)node[right]{$205$ keV} -- (0cm, 0.5em) node[left]{$\frac{7}{2}^+$};
      \draw[level, align=left] (1cm,2em)node[right]{$846$ keV, \\
        $\tau = 107(3)$~ns} -- (0cm, 2em) node[left]{$\frac{11}{2}^-$};

      \draw[transition] (0.5cm,2em) -- (0.5cm, 0.5em) node[right,
        midway] {$640$ keV};
      \draw[transition] (0.5cm,0.5em) -- (0.5cm, -0.5em);
    \end{tikzpicture}
    \end{sansmath}
    }}
  \end{overpic}
  \caption{Out-of-beam $\gamma$-ray energy spectra recorded by (a)
    HPGe and (b) LaBr$_3$ detectors following reactions of 48-MeV
    $^{12}$C on $^{98}$Mo.  The inset in (b) shows the region around
    the 640-keV transition of interest and indicates the peak and
    background regions used to project the time spectra in
    Fig.~\ref{fig:TAC}.}
  \label{fig:LaBr-spec}
\end{figure}

\begin{figure}
  \includegraphics[width=0.5\textwidth]{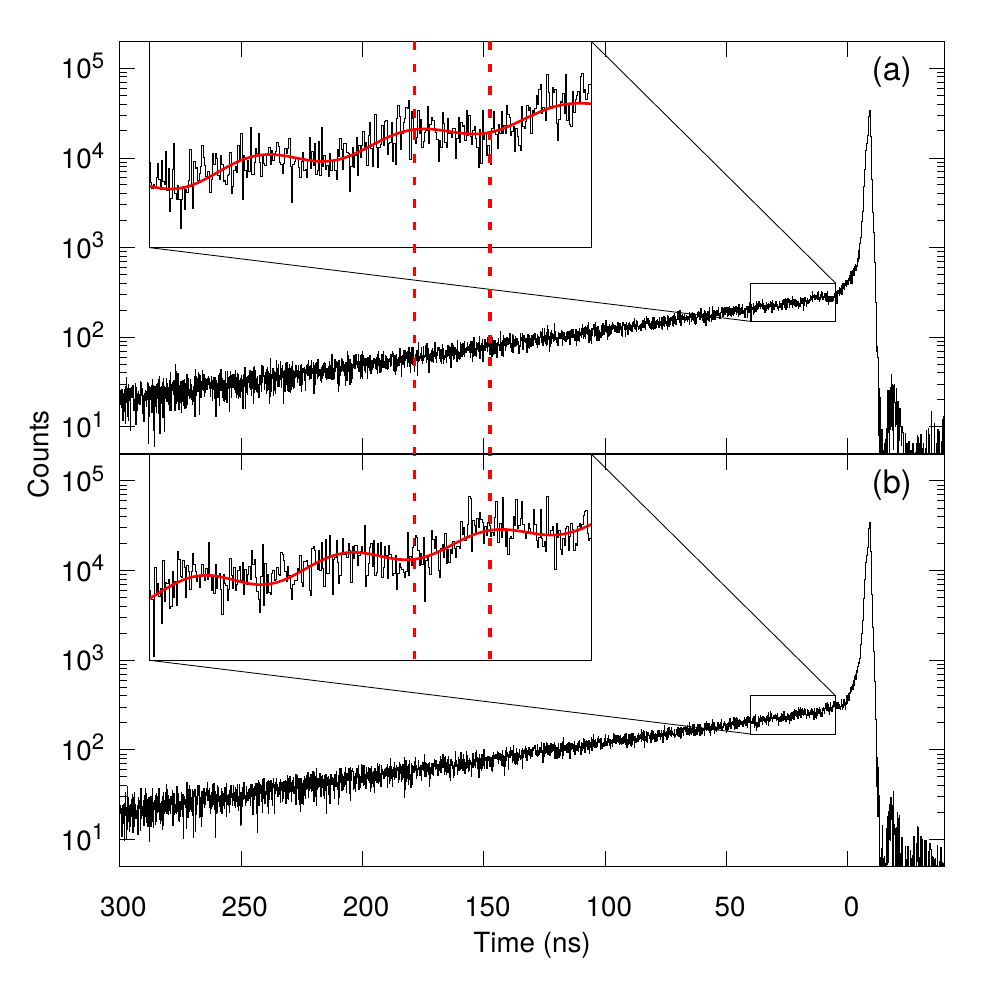}
  \caption{Time spectra for the 640-keV transition depopulating the
    $I^\pi=\frac{11}{2}^-$ isomer in $^{107}$Cd. (a): Group $N_1$ defined in
    Eq.~\ref{eq:n1}, (b): group $N_2$ defined in Eq.~\ref{eq:n2}.
    Fits to guide the eye show the oscillations in the two groups out of phase. The prompt
  peak is due to a prompt 632-keV transition in $^{106}$Cd which cannot be
  resolved from the 640-keV line by the LaBr$_3$ detectors.}
  \label{fig:TAC}
\end{figure}

Out-of-beam $\gamma$-ray energy spectra from HPGe and LaBr$_3$
detectors are shown in
Fig.~\ref{fig:LaBr-spec}.  Background-subtracted beam-$\gamma$ time spectra of the 640-keV transition are shown in
Fig.~\ref{fig:TAC}. The fitted mean life of $\tau~=~101(2)$~ns agrees
with previous measurements of $\tau=97(9)$~ns~\cite{Hagemann1974} and $\tau~=~111(10)$~ns~\cite{Bertschat1974},
and is within two standard deviations of the only other measurement,
$\tau~=~107(3)$~ns~\cite{Regan1995}. The fit accounts for $\approx 20\%$
feeding from the higher $I^\pi=\frac{21}{2}^+$, $\tau=79(6)$~ns isomer in
$^{107}$Cd~\cite{Hagemann1974, Regan1995}. The amount of feeding was
estimated from the HPGe energy spectrum, where the 520-keV transition
can be resolved. If the feeding from the
higher isomer is neglected, a value of $\tau=108(2)$~ns
results. The treatment of feeding might explain the discrepancy
between the present result and the lifetime reported
in Ref.~\cite{Regan1995}.

The data were histogrammed in two
groups to analyze the perturbed angular distribution. The first group
comprised Detectors 1 ($+45^\circ$) and 3 ($-135^\circ$)
when the field orientation was up, together with Detectors 2
($+135^\circ$) and 4 ($-45^\circ$) when the field orientation
was down.  The second group was the converse: Detectors 2 and 4 for
field up; and Detectors 1 and 3 for field down. Each detector within a
group is affected by the precession of the angular distribution in
the same way, whilst oscillations in the two groups should be
180$^\circ$ out of phase.  Figure~\ref{fig:TAC} shows time
spectra which display this behavior.
\subsection{Angular distributions}
\begin{figure}
  \includegraphics[width=0.5\textwidth]{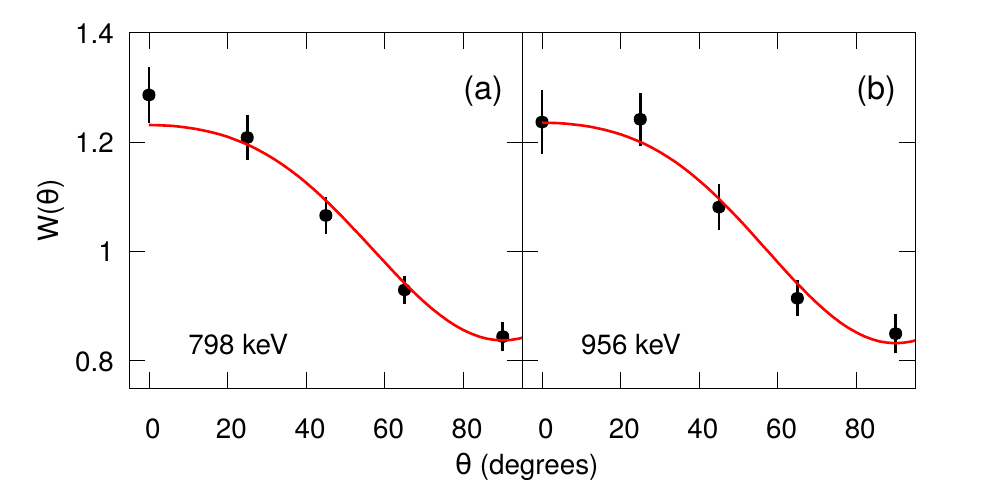}
  \caption{Angular distribution of the (a) 798-keV and (b) 956-keV transitions
  in $^{107}$Cd after population by the \mbox{$^{98}$Mo($^{12}$C, 3n)}
  reaction. These stretched $E2$ transitions
  both feed the 846-keV isomer of interest and give a good indication
  of its spin alignment.}
  \label{fig:angulardistr}
\end{figure}

Angular distributions were measured using a HPGe detector. The
anisotropy of the angular
distribution depopulating the isomer cannot be measured directly
because it loses its alignment during the long lifetime;
however,  decays of surrounding shorter-lived states give an indication of the initial
anisotropy of
the transition depopulating the isomer. Figure~\ref{fig:angulardistr} shows the angular
distributions of two of the transitions feeding the isomeric state.
Both transitions have pure $E2$ multipolarity; the 956-keV is a
transition between
$I^\pi=\frac{23}{2}^-$ and $I^\pi=\frac{19}{2}^-$ states, whilst the
798-keV is between $I^\pi=\frac{19}{2}^-$ and $I^\pi=\frac{15}{2}^-$
states. The solid lines represent fits to the
data with the spin alignment specified by a Gaussian distribution of
width $\sigma$~\cite{Stuchbery2002, Diamond1966, Yamazaki1967}. The
fit has two free parameters: $\sigma/I$ and a normalization factor.
For the 798-keV and 956-keV transitions the fitted values are
$\sigma/I=0.32(5)$ and $\sigma/I=0.30(6)$, respectively.
A value of $\sigma/I \approx 0.3$ is common for heavy-ion fusion-evaporation
reactions~\mbox{\cite{Stuchbery2002, Carpenter1990, Grau1974,
    Simms1974}}. The $\sigma/I$ parameter determines the anisotropy of
the angular distribution depopulating the $I^\pi=\frac{11}{2}^-$
isomer, and has an impact on the amplitude of the $R(t)$ functions
described in the following subsection.
\subsection{Ratio functions}
\label{sec:rf}
A ratio function formed from the time spectra shows
the precession of the angular distribution.  The standard form for two
detectors at $\pm45^\circ$ to the beam-axis is
\begin{equation}
  R(t)=\frac{N_1(t)-N_2(t)}{N_1(t)+N_2(t)}  \approx \frac{3A_{2}}{4 + A_{2}}\sin(2\omega_L t),
\end{equation}
\noindent where $N_{1}(t)$ ($N_{2}(t)$) denotes the number of counts
in the first (second) detector~\cite{Cerny1974}.  The approximate
expression applies when the unperturbed angular distribution can be
written as $W(\theta) = 1 + A_2 P_2(\cos(\theta))$, where $\theta$ is
the angle with respect to the beam axis, $P_2$ is a second order Legendre polynomial, and $A_2$
is an orientation parameter~\cite{Cerny1974, Christiansen1983}.  This form applies
to our setup if we assign
\begin{align}
  \label{eq:n1}
  N_1(t) &= N(+45^\circ)\uparrow + N(+135^\circ)\downarrow \\
  &\qquad \qquad + N(-135^\circ)\uparrow + N(-45^\circ)\downarrow,
    \notag
\end{align}
\noindent and
\begin{align}
  \label{eq:n2}
  N_2(t) &= N(+45^\circ)\downarrow + N(+135^\circ)\uparrow \\
  &\qquad \qquad + N(-135^\circ)\downarrow + N(-45^\circ)\uparrow, \notag
\end{align}
\noindent where $N(\theta)\uparrow$ ($N(\theta)\downarrow$) denotes a
detector at angle $\theta$ with respect to the beam axis with the
field up (down). A modified definition is used to subtract background:
\begin{equation}
  R(t) = \frac{T_1(t) - B_1(t) - T_2(t) + B_2(t)}{T_1(t)+T_2(t)-B_1(t)-B_2(t)},
\end{equation}
\noindent where $T_i(t)$ represents the total peak area (without background subtraction)
for the relevant group of detector/field direction combinations,
and $B_i(t)$ is the area of the background region, multiplied by a
scaling factor equal to the ratio of their widths (see
Fig.~\ref{fig:LaBr-spec}(b) inset).  In the present case, the background region shows no
evidence of precession effects, and $B_1(t) \approx B_2(t)$, thus
the form
\begin{equation}
  R(t) = \frac{T_1(t) - T_2(t)}{T_1(t)+T_2(t)-B_1(t)-B_2(t)}
\end{equation}
is used.
\begin{figure}
  \includegraphics[width=0.5\textwidth]{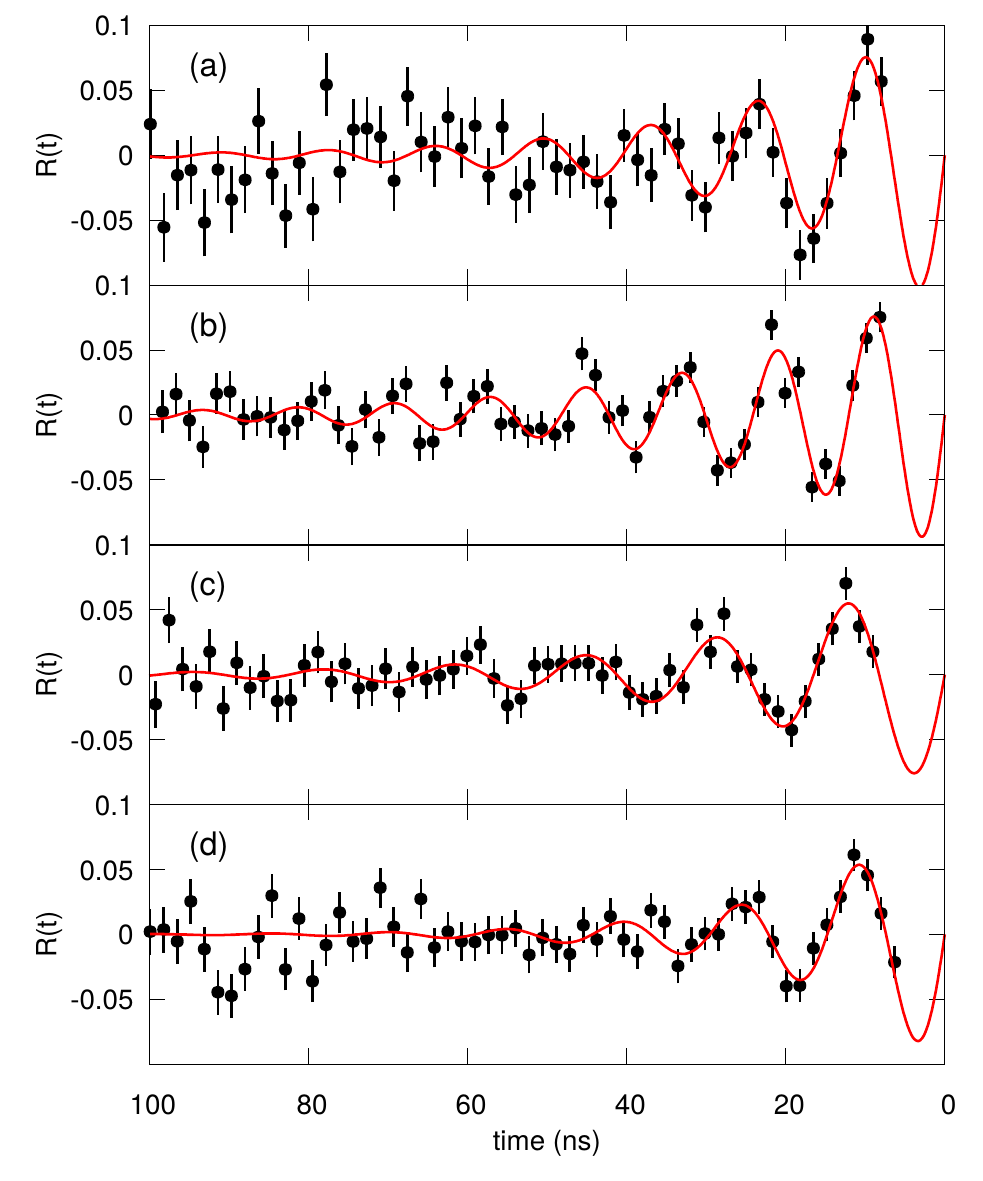}
  \caption{Ratio functions for $^{107}$Cd in gadolinium.
    (a)-(d) correspond to data sets I-IV in Table~\ref{tab:results}
    and in the text. Data sets are arranged in order of collection.
    The missing data near $t=0$ is on account of only using data well
    away from the (contaminant) prompt peak.}
  \label{fig:ratios}
\end{figure}
Ratio functions for a sequence of runs are shown in Fig.~\ref{fig:ratios}.  Each function
is formed from ${\sim}8$ hours of data collection. The ratio
function has been fitted with the form
\begin{equation}
  \label{eq:ratio}
  R(t)=C e^{-\nicefrac{t}{\kappa}}\sin(2\omega_Lt),
\end{equation}
\noindent where the exponential term is phenomenological, and commonly
used to account for decreasing alignment~\mbox{\cite{Mohanta2013, Mahnke1989}}. Table~\ref{tab:results} shows the fitted
parameters for the relevant data sets.
\begin{table}[]
  \centering
  \caption{Fitted parameters for ratio functions.}
  \label{tab:results}
\begin{ruledtabular}
\begin{tabular}{clll}
  Data Set & $\omega_L$ (Grad/s) & $\kappa$ (ns) & $C$ \\\hline
  I        & $-0.228(12)$ & $18(4)$ & $0.11(3)$\\ 
  II       & $-0.262(5)$ & $30(4)$ & $0.100(10)$\\ 
  III      & $-0.188(5)$ & $31(8)$ & $0.062(12)$\\
  IV       & $-0.209(8)$ & $19(6)$ & $0.07(2)$\\
\end{tabular}
\end{ruledtabular}
 \end{table}
\section{Analysis of Ratio Functions}\label{sec:analysis}
Two major features of the ratio functions shown in
Fig.~\ref{fig:ratios} are immediately apparent.  First, the fitted
amplitudes and frequencies vary significantly among data
sets. The reasons for this observation will be discussed in Section~\ref{sec:ard}.

Second, the amplitudes clearly
attenuate rapidly, with no more than three or four periods visible.
One possible
explanation for this attenuation is
that there are multiple oscillation frequencies present.  If
there were only
two or three distinct frequencies (equivalent to
the same number of field strengths), the frequency components would beat in and out of phase.  There is no
evidence of such behaviour occurring across the ${\sim}3$ mean lives observed, even
using autocorrelation or Fourier-transform analysis.  Thus a near
continuous distribution of fields is implied. This conclusion will be
discussed, and alternative
explanations of the observed $R(t)$ data will be explored, in Sections~\ref{sec:efg} and \ref{sec:dfs}.
\subsection{Accumulating Radiation Damage}
\label{sec:ard}
Both the initial amplitude ($C$) and the frequency ($\omega_L$) of the
$R(t)$ data
changed on macroscopic time scales. As well as being determined by
the anisotropy of the unperturbed angular distribution, $C$ is
dependent on the proportion of nuclei that are implanted into
field-free sites, $f$.  When $f$ is high, many nuclei decay without undergoing
precession, reducing the amplitude of the ratio function.

Data sets gathered later in the experiment show a decrease in both
$\omega_L$ and $C$, which can be attributed to accumulating radiation
damage to the gadolinium host. Thus, to relate the present
observations to the previous measurement of $g(10^+)$ in $^{110}$Cd,
it is important to match the level of accumulated beam dose.  The present work used
beam intensities up to an order of magnitude higher than those in the previous
$^{110}$Cd $g$-factor measurement~\cite{Regan1995}.  As a consequence, the equivalent
cumulative dose to the gadolinium host was reached before the end of
data set II (see Fig.~\ref{fig:ratios} and Table~\ref{tab:results}). For this reason we use only data sets I and II to
re-evaluate $g(10^+)$ in $^{110}$Cd in Section~\ref{sec:g-corr} below.
\subsection{Electric Field Gradients}
\label{sec:efg}
Along with the magnetic dipole interaction, the electric quadrupole
interaction associated with an electric field gradient (EFG) must be
considered in the case of a gadolinium host. The hexagonal close-packed (hcp)
crystal structure means that the quadrupole interactions do not cancel~\cite{Christiansen1983}.
The frequency ($\omega_Q$) associated with the EFG is given by
\begin{equation}
  \omega_Q = \frac{eQ}{4I(2I-1)\hbar}V_{zz},
\end{equation}
\noindent where $Q$ and $I$ are the electric quadrupole moment and
angular momentum of the nuclear state respectively, and $V_{zz}$ is the
$z$-component of the EFG~\cite{Christiansen1983}.

\begin{figure}
  \centering
  \includegraphics[width=0.5\textwidth]{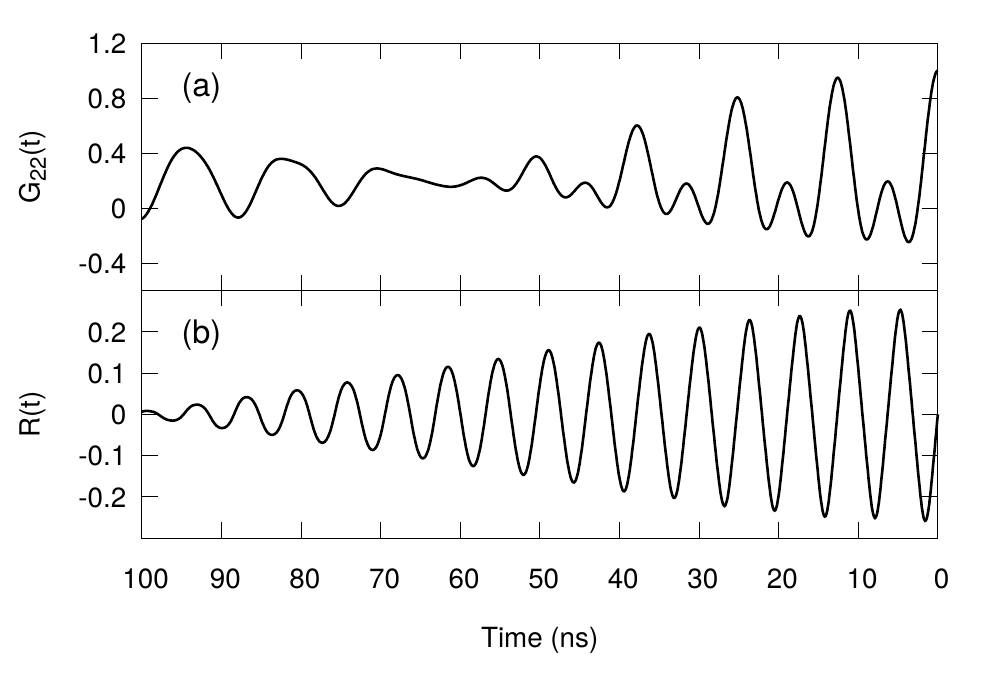}
  \caption{Simulation of (a) $G_{22}(t)$ and (b) $R(t)$ from combined
    magnetic and electric interactions for the $I^\pi=\frac{5}{2}^+$ state
    in $^{111}$Cd ($E_x=254$ keV, $\tau=121.9$~ns) in gadolinium. If
    the form of Eq.~\ref{eq:ratio} is assumed for $R(t)$, $\kappa
    \approx 61$~ns.  The $G_{22}(t)$ function corresponds to a
    polycrystalline source with no external field, whilst the $R(t)$
    function is applicable to the present experiment with a
    polarizing external field at $90^\circ$ to the plane of detection.}
  \label{fig:EBCd111}
\end{figure}

\begin{figure}
  \centering
  \includegraphics[width=0.5\textwidth]{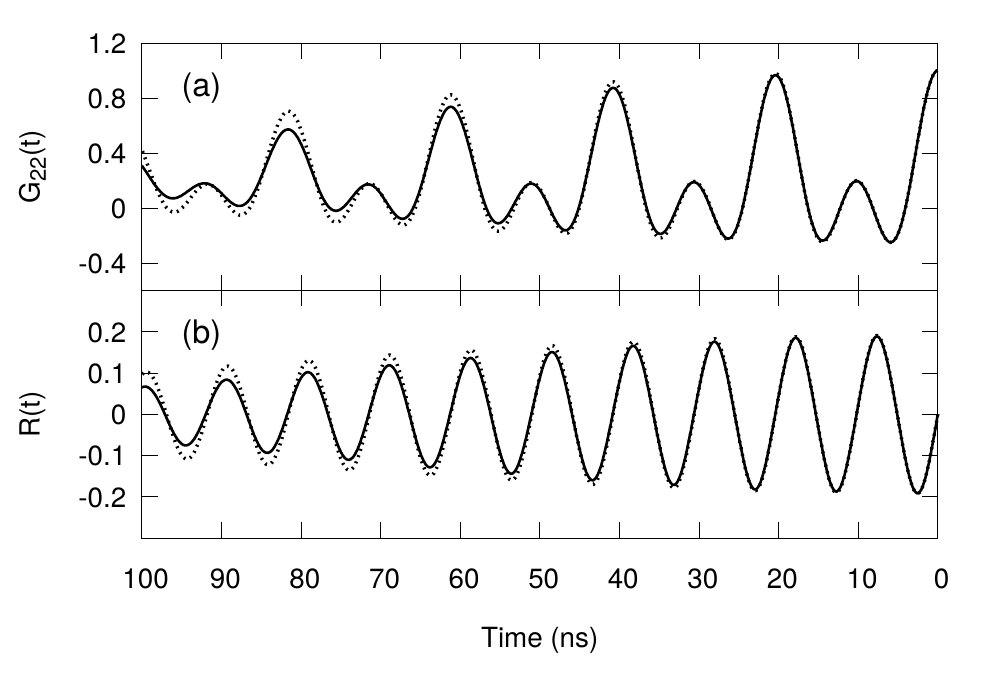}
  \caption{A simulation analogous to Fig.~\ref{fig:EBCd111}, showing
    (a) $G_{22}(t)$ and (b) $R(t)$ for the $E_x=846$ keV, $I^\pi=\frac{11}{2}^-$
    $^{107}$Cd state in gadolinium.  The solid line is for
    $\beta=30^\circ$, and the dashed for
    $\beta=90^\circ$.  If
    the form of Eq.~\ref{eq:ratio} is assumed for $R(t)$, $\kappa
    \approx 114$~ns. See also Fig.~\ref{fig:EBCd111} caption.}
  \label{fig:EBCd107}
\end{figure}

The combined electric-magnetic interaction has been examined thoroughly
for the $I^\pi=\frac{5}{2}_1^+$ state of $^{111}$Cd in
gadolinium by studying the $\gamma$-$\gamma$ angular correlations
after $^{111}$In decay~\cite{Bostrom1971, Forker1973, delaPresa2004}.
Each of these experiments used an amorphous sample with no polarizing
field. Thus the EFG and $B_{\rm hf}$ are randomly oriented;
however, there is a preferred angle ($\beta$) between $V_{zz}$ and $B_{\rm hf}$
for any individual gadolinium microcrystal~\mbox{\cite{Bostrom1970,
  Forker1973, Bostrom1971, delaPresa2004}}. In such experiments, the time-dependent angular
correlation function can be expressed as
\begin{equation}
  W(\theta, t) = 1 + A_{22}G_{22}(t) P_2(\cos\theta),
\end{equation}
\noindent where the $A_{22}$ coefficient is the $\gamma$-$\gamma$ angular correlation equivalent
of the $A_2$ discussed in Section~\ref{sec:rf}; the
$A_{44}$ term is neglected~\cite{Siegbahn1966}. These experiments
measured the perturbation factor $G_{22}(t)$ to obtain
the angle $\beta$, electric quadrupole frequency $\omega_Q$, and
magnetic dipole
frequency $\omega_L$. It should be noted that
$G_{22}(t)$ and $R(t)$ are fundamentally different observables that
apply to different experimental setups, although they
reflect the same physical phenomena. A direct comparison between the $G_{22}(t)$ functions obtained in the
off-line measurements, (Refs.~\cite{Bostrom1971, Forker1973,
  delaPresa2004}), and $R(t)$ of the present measurements that apply a
polarizing field, is not meaningful. However, $G_{22}(t)$ and the corresponding $R(t)$
applicable to our experiment resulting from the combined
electric-magnetic interaction can be evaluated. Examples of calculations for $^{111}$Cd
and $^{107}$Cd in gadolinium are shown in Fig.~\ref{fig:EBCd111}
and Fig.~\ref{fig:EBCd107}, respectively. The $G_{22}(t)$ calculations
assume a polycrystaline source with no external field applied, as in
Refs.~\cite{Bostrom1971, Forker1973, delaPresa2004}.  The calculated $R(t)$
functions, however, have an external polarizing magnetic
field applied perpendicular to the plane of the detectors as in the
present experimental setup.

Figure~\ref{fig:EBCd111} shows a simulation of
$G_{22}(t)$ and $R(t)$ from the combined magnetic and electric interactions for the $I^\pi=\frac{5}{2}^+$ state
in $^{111}$Cd ($E_x=254$ keV, $\tau~=~121.9$~ns) in gadolinium. A
conservative estimate of $V_{zz}~=~1.4~\times~10^{17}$~V/cm$^2$ is
used, with
$B_{\rm hf}=-34$~T, $\beta=30^\circ$~\cite{delaPresa2004}, $Q=0.77$~\cite{Raghavan1973}, and $g=-0.306$~\cite{Bertschat1974a}. The calculated
$G_{22}(t)$ matches that shown in Fig.~1 of Ref.~\cite{delaPresa2004}.  Note that the
ratio function for the same parameters shows very different behavior:
the attenuation due to
the electric quadrupole interaction is slower because the
magnetic interaction is held in the direction of the polarizing field.

Similarly,
Fig.~\ref{fig:EBCd107} shows both $G_{22}(t)$ and $R(t)$ for the $E_x=846$ keV, $I^\pi=\frac{11}{2}^-$
state of $^{107}$Cd in gadolinium.  The same $B_{\rm hf}$
and
$V_{zz}$ are used, with $Q=0.94$~\cite{Sprouse1978} and
$g=-0.189$~\cite{Raghavan1989}. Apart from the change in Larmor
frequency due to the change in $g$ factor, the striking difference in
$G_{22}(t)$ compared to Fig.~\ref{fig:EBCd111} stems from the change in nuclear spin.
It is clear from Fig.~\ref{fig:EBCd107}(b) that the electric quadrupole interaction is not nearly
strong enough to explain the decay of the ratio function as displayed
by the experimental data in Fig.~\ref{fig:ratios}.  In terms of the
effective decay constant, $\kappa$, the experimental data show
$\kappa \approx 18 \text{--} 30$~ns (Table~\ref{tab:results}), whereas
the evaluation of the effect of the EFG in Fig.~\ref{fig:EBCd107}
implies $\kappa \approx 114$~ns. It should be noted that the calculations are conservative (maximizing
the electric quadrupole interaction):
$V_{zz}=1.4 \times 10^{17}$~V/cm$^2$ was reported in Ref.~\cite{delaPresa2004}, whereas
two other measurements report smaller electric field gradients, $V_{zz} = 0.85 \times
10^{17}$~V/cm$^2$~\cite{Forker1973} and $V_{zz}=0.21 \times 10^{17}$~V/cm$^2$~\cite{
  Bostrom1971}. Also, in the case of our experimental geometry it is
highly likely the
$c$-axis of the hexagonal close packed structure (and so the EFG direction) is
perpendicular to the foil and hence parallel to the beam direction.  This is a
known property of cold-rolled and annealed gadolinium foils, which has
been observed in X-ray diffraction measurements and confirmed by
magnetization versus temperature curves~\cite{Stuchbery2006,
  Robinson1999}. With the $c$ axis along the beam direction,
$\beta=90^\circ$ and the effect of the EFG on the
ratio function is reduced.

In summary, it is evident that the effect of the EFG is not nearly
significant enough to account for the attenuation in the observed
$R(t)$ functions, and that EFG effects
can be neglected in further analysis.
\subsection{Distribution of field strengths}
\label{sec:dfs}
Another explanation for the attenuation of the ratio function is that a continuous distribution of
field strengths is present across a range of implantation sites instead
of a single, well-defined $B_{\rm hf}$. A ratio function can
be calculated using the angular distributions measured to set the
initial anisotropy, assuming alternative distributions of
field-strengths, and a field-free fraction $f$. The distribution parameters, such as the width and
average value, can then be fitted to the experimental data. The calculated ratio
functions were also attenuated to $0.87$ of the full amplitude to account
for the convolution of the beam pulse and time resolution of the
LaBr$_3$ detectors (${\sim} 2$~ns) with the $R(t)$ function. This
factor was calculated by evaluating the convolution of a sinusoid of
an appropriate frequency with
a Gaussian with FWHM of $2$~ns. The factor is not sensitive to small
changes in the frequency.

\begin{figure*}
  \centering
  \includegraphics[width=0.9\textwidth]{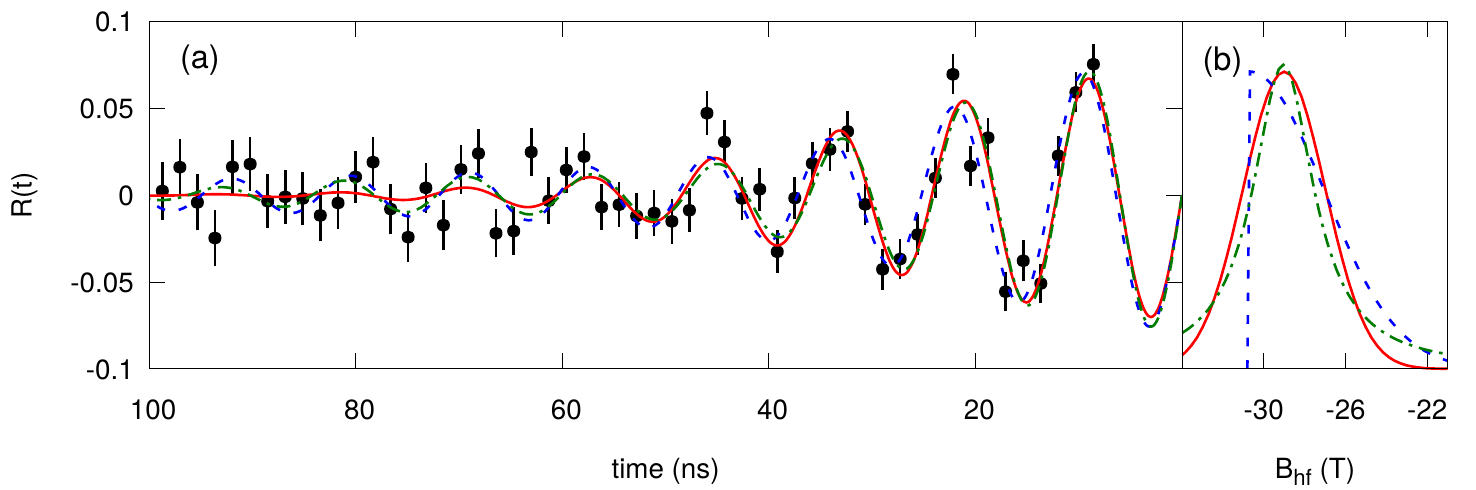}
  \caption{Ratio functions from different field
    distributions: red solid line,
    Gaussian; blue dashed line, Half-Gaussian; green dashed-dotted line, Lorentzian. (a): Fits of
    the ratio functions to data set II. (b): Field distributions,
    parameters specified in rows two to four in Table~\ref{tab:distr}.}
  \label{fig:distr}
\end{figure*}

A Gaussian distribution of hyperfine fields was found to reproduce the
observed $R(t)$ data, however as shown in Fig.~\ref{fig:distr} and
Table~\ref{tab:distr}, the ratio function is not sensitive to the
precise shape of the field distribution.  As evident from
Fig.~\ref{fig:distr} and the fit parameters in Table~\ref{tab:distr},
fits of equal quality were obtained with Gaussian, Lorentzian, and
Half-Gaussian field distributions for data set II (reduced $\chi^2 = 0.99, 0.99,$ and $1.00$, respectively). Along with the shape of the
field distribution, the fraction of nuclei on field-free sites ($f$)
was also fitted. For convenience, Gaussian field distributions were adopted for the re-analysis of the
$^{110}$Cd $g(10^+)$ measurement which follows.
\begin{table}[]
  \centering
  \caption{Field distribution parameters from fitting the $^{107}$Cd ratio
    function data (Fig.~\ref{fig:ratios}) in the process
    described in Section~\ref{sec:dfs}. $\Delta B_{\rm hf}$ is the
    FWHM of the distribution. }
   \label{tab:distr}
\begin{ruledtabular}
\begin{tabular}{ccclc}
  Data Set & $\overline{B_{\rm hf}}$ (T) & $\Delta B_{\rm hf}$ (T) &
                                                                     Distribution & $f$\\\hline
p  I   & $-24.3(4)$ & $6.6(8)$ & Gaussian      & $0.54(5)$ \\ 
 II   & $-29.0(2)$ & $4.7(4)$ & Gaussian      & $0.53(3)$ \\ 
  II  & $-29.0(2)$ & $3.6(3)$ & Lorentzian    & $0.45(3)$ \\ 
  II  & $-27.8(2)$ & $4.2(3)$ & Half-Gaussian & $0.52(3)$ \\ 
  III & $-20.7(3)$ & $4.4(5)$ & Gaussian      & $0.72(3)$ \\ 
  IV  & $-22.9(4)$ & $6.6(8)$ & Gaussian      & $0.70(4)$ \\
\end{tabular}
\end{ruledtabular}
 \end{table}
\section{Correcting the $\bm{g(10^+)}$ measurement}
\label{sec:g-corr}
The field-free fraction plays a dominant role in determining the
effective hyperfine-field strength for integral precession measurements like
the $g(10^+)$ measurement in $^{110}$Cd of Ref.~\cite{Regan1995}. The
usual expression for the integral perturbed angular distribution in
the case where there is a unique field and hence unique Larmor
frequency $\omega$, is:
\begin{equation}
  W(\theta)=\sum_k \frac{b_k}{\sqrt{1 + (k\omega \tau)^2}} \cos(k[\theta - \Delta\theta_k]),
\end{equation}
\noindent where $\Delta \theta_k$ are the solutions of $\tan(k\Delta
\theta_k) = k \omega \tau$, and the $b_k$ coefficients ($k = 0, 2, 4)$
are related to
the $A_k$ orientation parameters as given in Refs.~\cite{Cerny1974, Regan1995}.  With a distribution of fields, the expression
becomes
\begin{equation}
  \label{eq:wtheta}
  W(\theta)=\sum_{ki} \frac{p_i b_k}{\sqrt{1 + (k\omega_i \tau)^2}} \cos(k[\theta - \Delta\theta_{ki}]),
\end{equation}
\noindent where $\tan(k\Delta
\theta_{ki}) = k \omega_i \tau$, and $p_i$ is the fraction of nuclei
implanted into a site with field $B_i$, causing a precession at Larmor
frequency $\omega_i$.

\begin{figure}
  \includegraphics[width=0.45\textwidth]{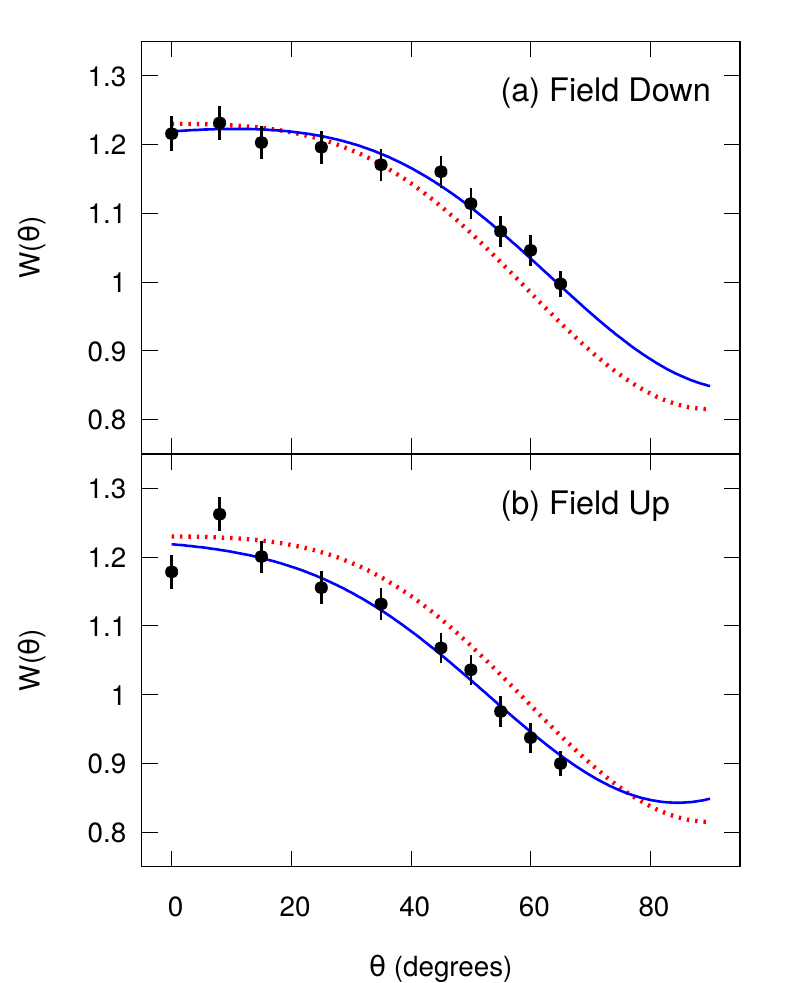}
  \caption{Fits to perturbed angular distribution data from
    Ref.~\cite{Regan1995}. A field distribution formed by
    taking a weighted average of the parameters extracted from
    data sets I and II was used (see the first two rows of
    Table~\ref{tab:distr}). The $g$ factor was
    varied to obtain the best fit.  The blue solid line is the perturbed angular
    distribution fit, the dotted red line is the unperturbed angular
    distribution.}

  \label{fig:pad}
\end{figure}

Equation~\ref{eq:wtheta} can be fitted to the original perturbed angular distribution data from
Ref.~\cite{Regan1995}, using the field distribution, including the
field-free fraction, taken from the
present $^{107}$Cd measurement.  A value of $\tau = 700(30)$~ps has
been adopted for the mean life of the
$I^\pi=10^+$ state in $^{110}$Cd~\cite{Kostov1998, Juutinen1994, Piiparinen1993,
  Harissopulos2001}. The distributions found in data sets I and II
(parameters on the first two rows of Table~\ref{tab:distr}) were
used separately and together. If the distribution formed by taking the
weighted average of the parameters from data sets I and II is used,
$g(10^+)=-0.29(16)$ is found from the resultant fit shown in
Fig.~\ref{fig:pad}. To assess the changes in effective field strength
on a macroscopic time scale, the $g$ factor was also evaluated based
only on data set I, giving $g(10^+) = -0.34$, and data set II alone
giving $g(10^+) = -0.28$, a difference of only 0.06 compared to the
0.16 uncertainty in the weighted average. These results show that the statistical
error from the original perturbed angular distribution measurement is
much more significant than the uncertainty from the variation in
fields between data sets I and II.
It is worth noting that the decay of the ratio function is of
little importance for the interpretation of the integral $g$-factor
measurement because the precession in the integral measurement
takes place in only the first few nanoseconds. The most significant impact on
the $g$-factor evaluation originated from the
fraction of nuclei implanted onto field-free sites.

In summary, we adopt the weighted average of data sets I and II to fit to the
perturbed angular distribution data from Ref.~\cite{Regan1995}. The
result is $g(10^+) = -0.29(16)$.
For a $\nu (h_{\nicefrac{11}{2}})^2$
configuration the Schmidt value is $g=-0.348$, or with quenching of
the spin $g$ factor to 70\% of the free nucleon value,
$g=-0.243$.
Thus the experimental value is consistent with the
$(h_{\nicefrac{11}{2}})^2$ neutron description of the state~\cite{Castel1990, Bohr1969, Frommgen2015, Yordanov2013}.

\section{Discussion}
\subsection{Comparison with previous work}
An in-beam time differential measurement of the hyperfine fields of Cd
in gadolinium has been reported here for the first time. The frequencies observed in Fig.~\ref{fig:ratios} imply hyperfine fields
close to, but slightly less than, what is expected from offline
measurements on Cd in gadolinium~\cite{Krane1983, Forker1973}. However, the
ratio functions observed in-beam attenuate more rapidly than expected
based on the off-line data.  This strong attenuation is
attributed to an effectively continuous distribution of hyperfine field values on
the implantation sites. The
width of the field-strength distribution is significantly larger than that
reported by
previous offline observations of hyperfine fields for Cd in
gadolinium~\cite{Forker1973}. However, the distribution widths (as a fraction of average field strength)
observed are comparable with
previous in-beam measurements on Ga, Ge, and As implanted into
gadolinium: between about $5\%$ and $18\%$~\cite{Raghavan1979, Raghavan1985}.
\subsection{Implications for other measurements}
The fraction of nuclei in field-free sites was significant. The
consequence is that the effective hyperfine field for in-beam integral
Perturbed Angular Correlation/Distriubtion measurements of Cd in gadolinium is much reduced compared to
offline measurements. Extracting the field-free fraction precisely
proved difficult as the
dependence of the initial amplitude of the ratio function on the width and shape of the
field-strength distribution makes the quantitative analysis complex
and multiplies uncertainties.

The precession frequency was also observed to vary on
macroscopic time scales.  This variation can only be attributed to a change in
the hyperfine field strength. The timing electronics were proven to be
stable because the mean-life measurements on subsets of the data were consistent throughout the
experiment. The changes in field-strength are most likely due to the
accumulation of radiation damage. In the case where the field strength
increased, we assume that the beam spot moved to an undamaged or less
damaged location on the target,
resulting in a temporary return to a higher average hyperfine-field
strength.

The observation that the later data sets have a much higher field-free
fraction is consistent with the suggestion that increasing
accumulated radiation damage is responsible. Unfortunately it was difficult to
replicate the accumulated radiation dose of the $g(10^+)$ measurement
precisely. However, the difference in effective fields and deduced $g$
factors for data sets I and II is small compared
to the uncertainties for the perturbed angular distribution data from the integral
$g$-factor measurement. Thus, despite the uncertainties in the evaluation of
the effective hyperfine field strength, it is now clear that the experimental $g(10^+)$ value is
consistent with that of the expected seniority-two $\nu h_{\nicefrac{11}{2}}$
configuration. A more
extensive $g$-factor experiment might involve measuring both the
time-integral $^{110}$Cd and the time-differential
$^{107}$Cd simultaneously, with a low beam current to avoid
accumulating radiation
damage. However, such experimental conditions are not easily
implemented.  An alternative host, which does not accumulate radiation
damage so severely, should be sought for future experiments.  For
example, it would be worthwhile to explore
the behaviour of Cd ions implanted into iron hosts. Previous
experiments implanting Ge into iron show no loss of alignment over
more than
${\sim}500$~ns~\cite{Raghavan1979,
  Raghavan1985}, in contrast to implantation into gadolinium where the
alignment is lost within ${\sim}100$~ns~\cite{Lee1988, Lee1991}.\\
\section{Conclusion}
LaBr$_3$ detectors have been applied to the in-beam TDPAD technique and their effectiveness has been demonstrated by
the measurement of a frequency that proved too fast to resolve with HPGe
detectors.  Future applications of LaBr$_3$ detectors to
measure precessions with periods of ${\sim} 5$~ns in-beam are feasible.

There are, however, unanswered questions about gadolinium as a
ferromagnetic host for in-beam $g$-factor measurements of this type.  Whether the
behavior observed here (significant distribution of fields, variation of field
strength on macroscopic time scales) is typical of gadolinium as a
host in general, or specific to the case of Cd in gadolinium studied
here, remains to be investigated more thoroughly.

Despite these uncertainties, it is clear that the previous $g(10^+)$ measurement in
$^{110}$Cd was based on an incorrect value for the effective hyperfine field.
With the field corrected from the present study, the $g$ factor
becomes consistent with the
theoretical understanding that the $I^\pi=10^+$ state is associated with a
seniority-two $\nu h_{\nicefrac{11}{2}}$ configuration. This example demonstrates the value of
time-differential techniques as a complimentary tool to validate or calibrate
time-integral $g$-factor measurements.

\begin{acknowledgments}
The authors are grateful to the academic and technical staff of the
Department of Nuclear Physics (Australian National University) and the
Heavy Ion Accelerator Facility for their continued support. Thanks are
due to J.~Heighway for assistance in making the target, and to S.~Battisson for making the mu-metal shielding. This research was
supported in part by the Australian Research Council grant numbers
DP120101417, DP130104176, DP140102986, DP170103317, DP170101673,
LE150100064 and FT100100991, and by The Australian National University
Major Equipment Committee Grant no. 15MEC14.
T.J.G., A.A., B.J.C., J.T.H.D., and M.S.G. acknowledge support of the
Australian Government Research Training Program.

\end{acknowledgments}

\bibliographystyle{apsrev}
\bibliography{Cd107}

\end{document}